# A Contextual Topic Modeling and Content Analysis of Iranian laws and Regulations

Zahra Hemmat[1*], Mohammad Mehraeen[1], Rahmatolloah Fattahi[2]

[1] Department of Management, Faculty of Economics and Administrative Sciences, Ferdowsi University

of Mashhad, Mashhad, Iran,

Hemmat@mail.um.ac.ir,

Mehraeen@um.ac.ir,

[2] Department of Library and Information Science, Ferdowsi University of Mashhad, Mashhad, Iran,

fattahi@ferdowsi.um.ac.ir,

**Abstract**

A constitution is the highest legal document of a country and serves as a guide for the establishment of other laws. The constitution defines the political principles, structure, hierarchy, position, and limits of the political power of a country's government. It determines and guarantees the rights of citizens. This study aimed at topic modeling of Iranian laws. As part of this research, 11760 laws were collected from the Dotic website. Then, topic modeling was conducted on the title and content of the regularizations using LDA. Data analysis with topic modeling led to the identification of 10 topics including Economic, Customs, Housing and Urban Development, Agriculture, Insurance, Legal and judicial, Cultural, Information Technology, Political, and Government. The largest topic, Economic, accounts for 29% of regulations, while the smallest are Political and Government, accounting for 2%. This research utilizes a topic modeling method in exploring law texts and identifying trends in regularizations from 2016-2023. In this study, it was found that regularizations constitute a significant percentage of law, most of which are related to economics and customs. Cultural regularizations have increased in 2023. It can be concluded any law enacted each year can reflect society's conditions and legislators' top concerns.

## 1. Introduction

The constitution consists of a set of legal and political rules that are binding on all citizens and institutions. It is concerned with the structure and operation of government institutions, political principles, and citizens' rights. Public legitimacy underpins the constitution [1]. Legislative texts differ in aspects such as structure, vocabulary, and ambiguity from a newspaper article or scientific paper [2, 3] which makes them difficult to understand. Legal text is typically arranged in a number of small units and paragraphs that are only connected through overt linguistic means to a limited extent. In addition, legal texts may only contain normative statements and may not contain commentary, explanations, or metalinguistic remarks that are commonly used in ordinary language to facilitate understanding [4].

In the legal field, automated systems are becoming increasingly common as a means of managing rapid changes [5]. An increasing amount of online legal information has made legal text processing an important research area [3]. It is helpful to take a look at the core topics within the domain in order to identify salient terms that appear across various documents [5]. The topic modeling is a type of machine learning used to identify the primary themes throughout a collection of documents [6]. Topic modeling can be used to classify documents and identify word-based patterns [7]. An advantage of topic modeling is that it is unsupervised and does not require prior annotation [8]. Using topic modeling provides a framework for organizing, recognizing, and summarizing large amounts of textual information, in addition to providing an opportunity to discover latent topics [9].

Legal text analysis is becoming increasingly popular in research. Topic modeling techniques are commonly used to discover clusters from legal corpora such as constitutions, statutes, and parliamentary records [10, 11]. O'Neill et al. [5] conducted a research study regarding topic modeling of United Kingdom legislative texts comprising 41,518 documents between 2000 and 2016. For topic modeling, Saffron (a software for constructing a model-free topic hierarchy), Non-Negative Matrix Factorization (NMF) [12], Latent Semantic Analysis (LSA) [13], Latent Dirichlet Analysis (LDA) [14], and Hierarchical Dirichlet process (HDP) [15] were used. Soh and Chai [16] compared the effectiveness of different machine learning algorithms for categorizing verdicts into legal categories. A dataset of 6,227 Singapore Supreme Court rulings was used to compare natural language processing techniques such as topic modeling, word embeddings, and language models with statistical models for legal documents. Topic modeling was performed using LSA, LinSVMs, and LDA.

Topic modeling was used by Wendel et al. [17] to analyze the case law of the German Federal Constitutional Court. LDA was used to categorize 9,267 decisions published between 1951 and 2017 and words associated with certain areas of law are identified. Using LDA, Dalwadi [18] extracted common themes and topics across North Carolina's 3,01,328 session laws from 1867 to 1968. This article identifies 28 topics throughout the corpus of laws, analyzes differences between topics over time, and identifies racially biased legislation topics. Remmits [19] used LDA to extract case law topics from Dutch supreme court decisions. A total of 50 topics were extracted and a human evaluation was also conducted. According to the results, only half of the topics were rated as coherent by the subjects, and for most documents, the main topic was not identified by the model. The human evaluation of results indicates that domain experts may evaluate topics differently from non-domain experts.

In a study conducted by O'Halloran et al. [20], machine learning and natural language processing were used to analyze the United States Government's regulation of the banking and financial services sector. A LDA topic modeling technique was applied to financial regulation laws enacted between 1950 and 2010. Silveira et al. [21] utilized BERTopic modeling for the topic modeling of legal documents. Data were collected from Cornell Legal Information (Cornell LII)'s repository of Historic US Supreme Court Decisions and 314 cases were randomly selected for evaluation. Raghuveer [22] clustered Indian legal judgments based on topics derived from LDA. For clustering, cosine similarity was used to measure similarity between documents.

Rather than searching for document similarities in the corpus vocabulary space, George et al. [23] used LDA and LSI to identify document similarities in the topic space. In addition, they developed a SVM model to classify the documents. The proposed model was evaluated on the TREC-2010 Legal Track query dataset. Using topic segmentation, Lu et al. [24] propose an approach to cluster legal documents. The documents include "judicial opinions, statutes, regulations, administrative materials and analytical documents". Topic modeling is performed using metadata and SVM ranker. To evaluate the quality of the topics extracted from the algorithms, the opinions of legal experts were used as a baseline. In the proposed approach by Luz De Araujo and De Campos [25], LDA is used to model 45,532 lawsuits filed in the Brazilian Supreme Court. A semantic analysis of subjects revealed that models with 10 and 30 topics could capture some of the legal issues discussed by the court.

Ash and Chen [26] examined judges' patterns on a set of United States Supreme Court decisions using document embeddings that distinguish courts, time, and legal topics. Carter et al. study [11] analyzed 7476 decisions from the High Court of Australia ('HCA') spanning from 1903 to 2015. The authors used LDA to generate topics from a corpus and reported 10 and 50 extracted topics respectively. A study by Li et al. [27] examined China's land policy from 1998 to 2018. In order to extract temporal evolution of policy and spatial differentiation of policies, the authors applied LDA techniques to 20,000 policy documents at different levels of government. In Bourgeois et al. [28], the French Parliamentary Debates of two decades (1881-1899) during the Third Republic (1870-1940) were analyzed using natural language processing techniques and topic modeling.

To represent contextualized words at the token level, Thompson and Mimno [29] used BERT [30], GPT-2 [31], and RoBERTa [32]. Based on these contextualized representations, the k-means algorithm generates topics for legal opinions presented by the Supreme Court of the United States. Chalkidis et al. [33] introduced the LEGAL-BERT model, a family of BERT for the legal domain. LEGAL-BERT has been pre-trained using EU and UK legislation, European Court of Justice cases, European Court of Human Rights cases, as well as US court cases and US contracts. To study legal research on artificial intelligence, Rosca et al. [34] applied LDA topic modeling to 3931 journal articles. Based on the results, 32 meaningful topics have been identified and legal research on artificial intelligence has significantly increased since 2016.

The current study utilized the LDA algorithm to determine the trend of Iranian law based on the textual data collected from the Dotic website [35]. Dotic is the Iranian portal for law and regularizations. Proposed method identifies the most important subjects and topics based on TF-IDF weights. This research has several contributions and is novel in several ways. The law data provided here are derived from Dotic, and have been gathered using coding. Consequently, a dataset of Iranian law and regularizations is collected. There is no research on text mining in order to discover Iranian law topics using the topic modeling method and the LDA algorithm. This is the first study in this field.

## 2. Methodology

In this study 11760 laws collected from July 2016 to July 2023 are examined. Data extraction was performed using two Python libraries, BeautifulSoup and Request. The dataset contains a variety of features, such as title, content, lead (a summary of law), tags, classes, types, categories (law source organization), and date. Translations of samples of collected laws are shown in Table 1. A variety of legal texts are presented in dataset, including news, drafts, votes, plans, laws, bills, parliament deliberations, regulations, and opinions. Topic modeling was conducted by LDA using Python programming language only for regulations.

Table 1: Sample of collected laws.

| Title | Content | Tags | Classes | Date | Type | Categories |
|---|---|---|---|---|---|---|
| Deliberations summary of the public session of the Islamic Council on Tuesday, July 6, 1402. | Deliberations summary of the public session of Tuesday, July 6, 1402 of the Islamic council was published. The main discussions of this meeting are as follows: 1- The report of the judicial and legal commission on the performance of the bar associations and the union of bar associations of Iran - in the implementation of Article 212 of the internal regulations law 2- Reading the report of the commission on Article 90 of the constitution regarding air pollution. | Air pollution, 90th principle commission, Water resources, Water resources of Iran, Bar association | The deliberations of the Islamic Council, The Islamic Council | Saturday, July 10, 1402 | Parliament deliberations | Islamic council |
| | 3- Announcing the receipt of a project item:<br>- Plan to amend articles of Iran's aquatic resources protection and exploitation law.<br>4- written reminders of parliamentarians to executive officials of the country. | | | | | |
| Seventh five-year development plan bill (1402-1406). | The bill of the seventh five-year development plan (1402-1406) was sent to the Islamic Council by the president on 03/28/1402 for legal formalities. | Development, The seventh development plan, The development plan, The country's development plan, The economic, Social and cultural development plan | Bills news, Economic group, Cabinet | Monday, June 29, 1402 | Bill | The Council of Ministers |
| The opinion of Islamic Council Chairman regarding the approval of the Supreme Council for the implementation of the general policies of Article 44 of the Constitution on the issue of determining the mechanism for the transfer of equity shares to qualified individuals. | The opinion of Islamic Council Chairman regarding the approval of the Supreme Council for the implementation of the general policies of Article 44 of the Constitution on the issue of determining the mechanism for the transfer of equity shares to qualified persons was announced by letter No. 20146/HB on 03/17/1402. | Shares, Justice shares | News, Legal opinions, Economic group, Islamic Council Chairman, All executive organizations | Monday, June 22, 1402 | Opinion | Islamic Council Chairman |
| The Executive Regulations of Clause A, Note 6 of Article One of the Budget Law of 1402 for the whole country. | The approval letter of the Council of Ministers meeting dated 04/04/1402 regarding the "executive regulation of paragraph A of note 6 of the single article | Budget, 1402 budget, 1402 budget law, 1402 budget law of the whole country, Villa garden, Authorized villa garden, | Regulations news, Financial and Economic Group, Cabinet of Ministers, Ministry of Economic Affairs and Finance, | Sunday, July 11, 1402 | Regulation | The Council of Ministers |

|  | of the budget law of the year 1402 of the whole country" was notified by the first vice president in letter number 59606 dated 04/07/1402. | Unauthorized villa garden | Ministry of Justice, Ministry of Agricultural Jihad, Ministry of Roads and Urban Development, Ministry of Interior, Ministry of Industry, Mining and Trade, Organization of Records and Real Estate of the country |  |  |  |
|---|---|---|---|---|---|---|
| Designing a list of invalid rulings in the field of business. | The draft of the list of invalid rulings in the field of trade was received in the public session dated 03/22/1402 of the Islamic Council. | Trade, Invalid rulings in the field of trade | News of projects, Economic group, Islamic Council | Sunday, June 28, 1402 | Plan | Islamic council |

## 2.1. Preprocessing

Data preprocessing is essential, which involves transforming raw data into a format that can be understood by machines. Unstructured text data can be preprocessed to extract non-trivial information [36]. In order to prepare the data for topic modeling, the following steps were required to preprocess the data:

- Checking for empty records. The dataset did not contain any NULL values.
- Punctuation Removal: Unstructured documents have numerous punctuations such as apostrophes, commas, and so on, which are not understood by the machine [37].
- Tokenization: A stream of text is broken down into words, phrases, symbols, or other meaningful elements, which are referred to as tokens [36].
- Stop Word Removal: In textual documents, these words are often used to join words together, but they don't add much value to the content. There is a list of Persian stop words provided by [38], which has been used in this study.
- Lemmatization: It involves breaking down words into their root [39].

The data was preprocessed using the Hazm and Parsivar Python libraries.

Next, vectorization is performed using TF-IDF, which converts the texts into numbers. Through the TF-IDF, we are able to determine the amount of information that a word provides about the context. Term frequency refers to how frequently a term appears in a document, and how relevant the term is to the document. As a result of using the TF-IDF, relativeness of the words in the document is determined, and the more informative words rule out the frequent word [40].

*2.2. Topic modeling*

Topic modeling identifies words within a set of documents and discovers patterns within them, providing an opportunity for clustering of words or comparative phrases [7]. In this step, the input from the TF-IDF is used by the LDA algorithm, which is one of the most popular and classic methods for topic modeling [41]. Based on the LDA approach, documents are considered to be generated as the result of randomized mixtures of hidden topics, which can be viewed as probability distributions over words [42]. As part of LDA's topic modeling methodology, each document is considered a collection of topics in a certain proportion. In addition, each topic is composed of keywords, again in a specific proportion. The algorithm will rearrange the distribution of topics within the documents and keywords within the topics based on the number of topics provided. This will result in a good balance in the distribution of topics and keywords [43].

This research project was implemented using Gensim, Pandas, Numpy, and Sklearn. An excellent implementation of Latent Dirichlet Allocation (LDA) is available in Gensim library [43]. In the Pandas library, the Series and DataFrame are the main structures that provide a way of organizing dissimilar types of data into a single data structure and enabling the application of methods or functions to the entire data set [44]. NumPy is the primary array library for the Python [45]. Arrays in NumPy enable a wide range of scientific computations by efficiently storing and accessing multidimensional arrays (also known as tensors) [46]. The Sklearn (Scikit-learn) library in Python is one of the most powerful and useful libraries for machine learning and statistical modeling [47].

## 3. Results

We retrieved 11760 laws from Dotic, categorized into different types. Figure 1 shows the evolution of types of laws (2016-2023). There is an increase in the number of laws issued from 2016 to 2018 except for those related to news and voting. From 2018 to 2019, the number of laws being made decreased, then increased until 2021. Following that, the number of laws is reduced. The majority of the laws are issued in 2021. Regulations are the most common type of law.

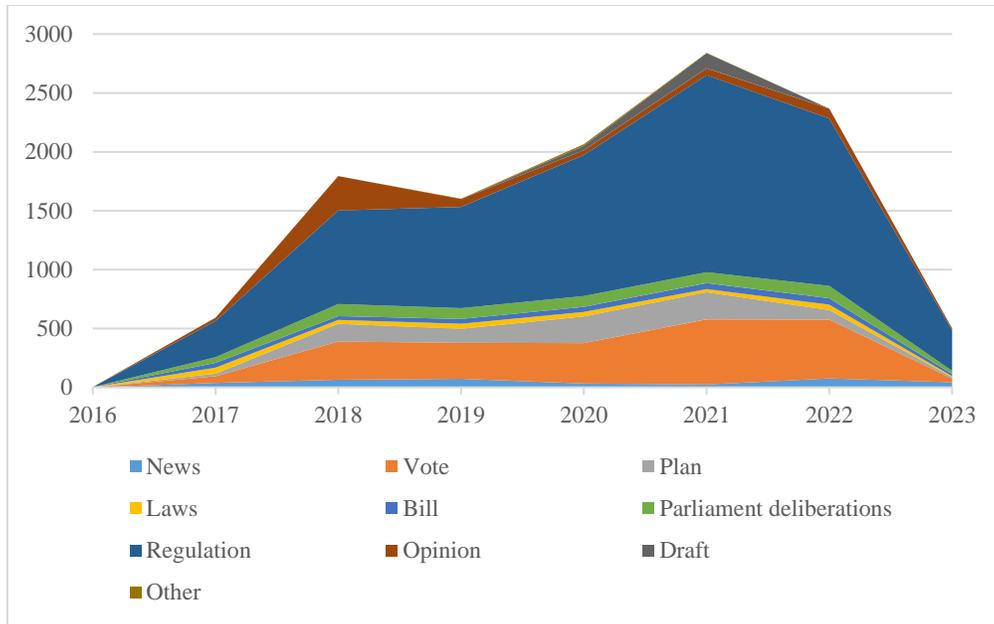

Figure 1: The evolution of laws over time.

Each type of law has its own characteristics, resulting in a variation in the size of text in the title and in the content. In Figure 2, each law is presented in a ratio of title size to content size. For example, the title of a news is usually brief, whereas the content is usually far more extensive. The title of Parliament approvals tends to be longer than news titles, as the longer the title is, the more understandable the approvals become, and they often contain numbers as well as parentheses.

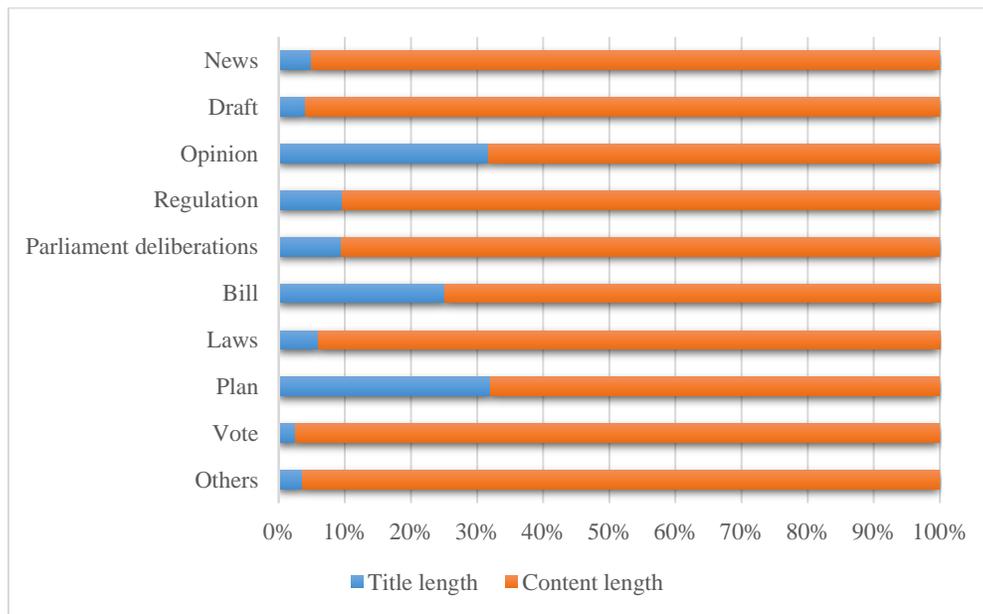

Figure 2: Title length to content length ratio.

The difference between law and regulation based on the definition provided by the Dotic website should be noted. Law refers to the decisions that create rights and duties for individuals (real and legal), or for the three powers (legislative, executive, and judicial) and other institutions, from which other regulations are derived. Regulations are the decisions that determine the method for the implementation of one or more laws, the regulation of administrative organizations, and other decisions that are made and implemented by the competent authorities.

Topic modelling was conducted using the title and content of regulations, since these features contain the most useful information that enhance the analysis. A total of 6599 records of regulations were analyzed using the LDA algorithms. Regulations make up a greater percentage of the law, and their content is more detailed and targeted, which makes their text more suitable for processing. Ten main topics were identified from these records. The frequency of each topic is shown in Figure 3. The most regulations relate to economics (29%), customs (26%), and housing and urban development (16%).

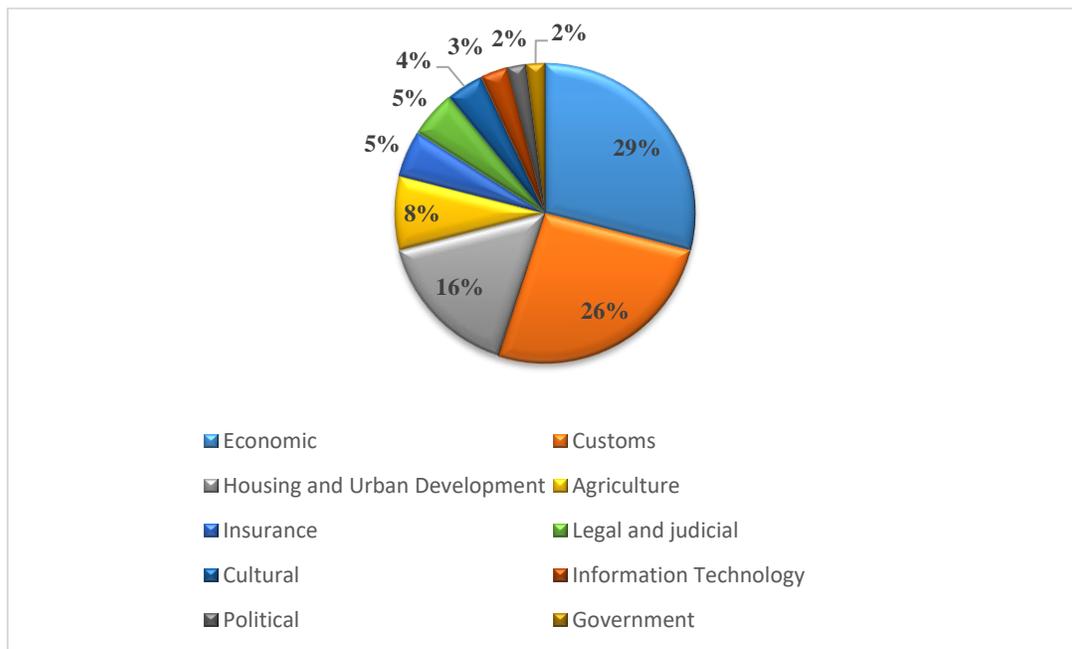

Figure 3: The frequency of topics in regulations.

According to Figure 4, all topics has minimum law in 2016 which is consistence with Figure 1. A total of 47% of regulations in 2017 related to political issues, and 35% in 2018 related to customs issues. The Insurance topic has the highest number of regulations in 2019 with 49%. The most important topic in 2020 was Legal and judicial, which accounted for 56% of the regulations. The most regulations in 2021, 2022, and 2023 will come from Economic (47%), Agriculture (59%) and Cultural (53%), which is completely consistent with the conditions of the Iran society.

A number of events occurred in 2021, including the presidential election and the change of government, protests of farmers due to the drought and their problems, as well as the suspension of negotiations to revive the JCPOA. In addition to affecting economic conditions of the country, these problems have resulted in an increase in prices, which could be a factor in the increase in economic regulations.

Drought and rising temperatures created many problems for farmers in 2022. Agricultural products are also damaged by these droughts, which can lead to food shortages in the country. Agricultural regulations may be increased as a result of these problems. Also, these regulations may have been effected by protests that took place in 2021 in Khuzestan and Isfahan.

Guidance patrols were reinstated with a large extent in various cities at the middle of 2022, resulting in widespread protests across the country. This led to the hijab becoming one of the most important issues in the country. The presence of the most regulations for the cultural topic in 2023 can be attributed to these cases.

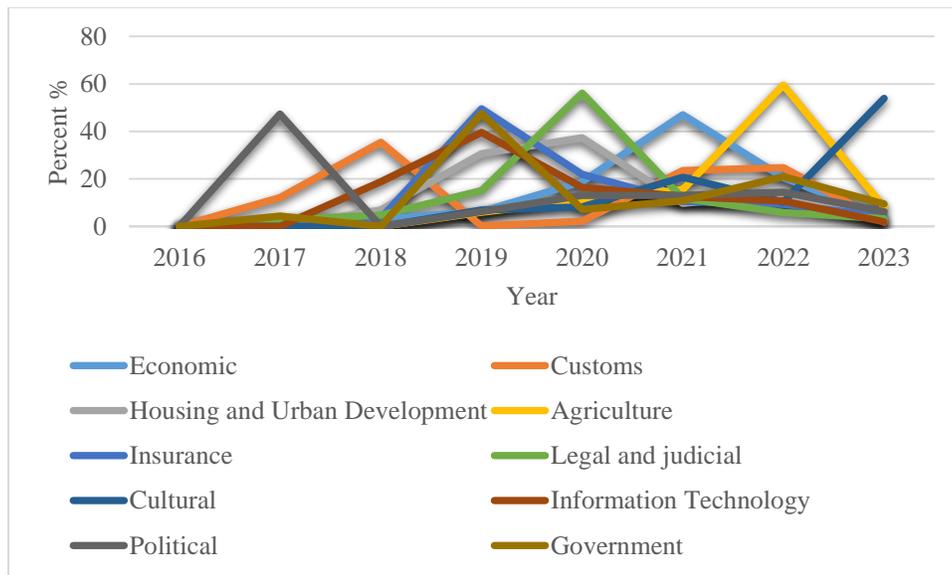

Figure 4: The percent of regulations per year for each topic.

Figure 5 illustrates the top word clouds in the ten identified topics, which are explained in the following.

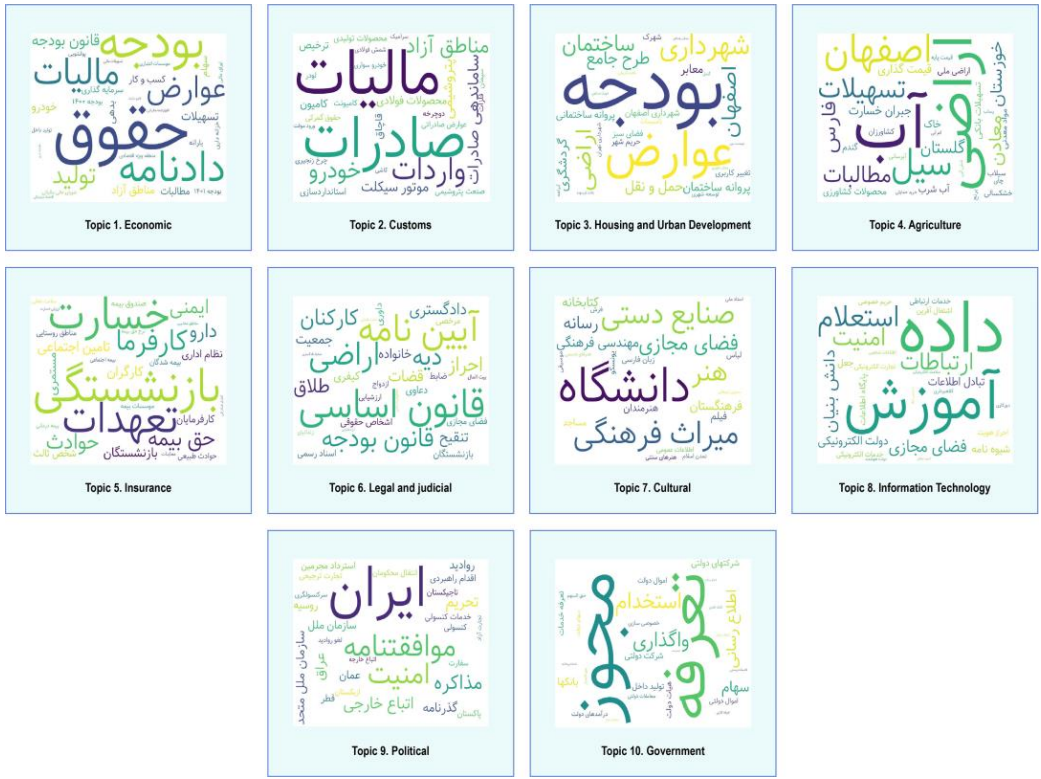

Figure 5: Topics identified by combining LDA+TF_IDF algorithms.

## 3.1. Topic 1. Economic

Economic structure plays an important role in determining economic growth, energy demand, and the environmental footprint of countries [48]. An increase in domestic production leads to economic growth and increase the budget. Governments plan their spending based on the budget each year. But in recent years, the Iran government has run a severe budget deficit as a result of economic problems. The government debt has risen, businesses have reduced many of their products, and domestic production has decreased as well. Many governments use the budget deficit policy to eliminate temporary revenue shortages and to achieve targeted economic growth. Recent years, the Iranian government has received some parts of the budget deficit from a variety of sources in order to address different economic problems. An important source of deficit financing is foreign borrowing, which should be used to increase employment [49]. Legislators need to revise the laws in order to address these issues for achieving economic growth.

## 3.2. Topic 2. Customs

Customs is a term used internationally. According to the Customs Co-operation Council (C.C.C.), customs is an organization tasked with enforcing customs laws, collecting import and export duties, as well as importing, transiting, and exporting goods [50]. Based on resources and facilities available, countries design and implement a variety of programs in order to achieve their development goals [51]. One of the key pillars of trade facilitation and growth in countries, especially developed ones, are free zones, which also play a crucial role in increasing exports and trade volumes. A free zone's establishment and continued operations will be supported by the facilitation of laws and structural protections [52].

Export and import policies should be developed in a manner that results in exports in excess of a country's needs, and imports in excess of the country's needs. For this reason, legislators are responsible for specifying duty laws according to the country's needs for various products, including steel, petrochemicals, cement, metals, and motorcycles and bicycles, so that society's needs can be met and smuggling is prevented. In its most basic form, smuggling is the illegal importation of goods from one jurisdiction to another [53]. A transaction that is illegal occurs in order to avoid legal taxation and duties [54]. It is necessary to set laws and penalties in order to prevent smuggling.

*3.3. Topic 3. Housing and Urban Development*

Today, urban development is growing at such a rapid rate that the earth appears to be becoming an urban environment [55]. It has become increasingly common for municipalities to rely on the financial resources of property developers and investors to implement comprehensive urban development plans. Several studies have been conducted to analyze the different methods used to charge developers so that the expenses of municipalities can be funded [56].

Most countries are transitioning from relying exclusively on traditional sources of funding (e.g. property taxes, duties, and intergovernmental transfers) to alternative sources of revenue generation. It is likely that property taxes will not be sufficient to cover the expenditures of municipalities in most cities due to a reduction in intergovernmental transfers. As a result, municipalities seek alternative ways to access private capital in order to make up for this deficit [57].

A major source of municipal income is the issuance of building and construction permits. There must be laws governing the issuance of permits and the collection of duties. Infrastructure costs, such as roads, can be covered by these charges, as well as the costs of public services, such as schools and parks.

*3.4. Topic 4. Agriculture*

The impacts of climate change are already being observed around the world, including an increase in the frequency of weather events such as droughts, floods, heat waves, wildfires, rising sea levels, and reductions in biodiversity. As a result of these events, people's lives can be negatively impacted in a number of ways, including health problems, economic losses, loss of labor productivity, damage to housing and critical infrastructures, as well as social disruptions [58].

As a result of climate change, droughts and water scarcity have caused extensive damage to agriculture. It is important to note that each province has different capacities in terms of agriculture and water resources. Therefore, laws can be enacted to specify the types of products that are allowed for agriculture in each province. The provinces of Khuzestan, Isfahan, and Fars are experiencing water scarcity, and farmers have had to deal with many challenges as a result. To support them, legislation must be enacted, which can support pricing and provide banking services. In order to solve the problem of water scarcity, there should be laws governing water saving, water efficiency, water pricing policy, cost-effective alternative solutions, and water hierarchy.

*3.5. Topic 5. Insurance*

The purpose of insurance is to provide financial protection against any type of risk. As part of the insurance contract, the insurer agrees to take on the risk of an insured entity against future events in exchange for monetary compensation [59]. Various types of insurance policies are available depending on the type of risk involved, including life insurance, travel insurance, health insurance, and property insurance. Laws must be specified for each type of insurance, so that insurance companies can correctly assess the amount of damage and beneficiaries can receive their compensation when needed.

*3.6. Topic 6. Legal and judicial*

There are many areas of government, politics, and civil rights that are covered in the constitution, which is the highest legal document of a country. To provide fair judicial decisions in criminal and legal cases, judges refer to the constitution and its regulations as official documents of the country. A large number of issues relating to people's social lives are covered in the constitution, including family, marriage, divorce, as well as the conditions of employees and retirees. Since there are many types of lawsuits, it is necessary to specify the applicable law for each.

*3.7. Topic 7. Cultural*

The concept of culture can be described as multifaceted, encompassing behaviors, values, and attitudes that are intrinsic to a particular group of people [60]. There are many components of culture, including knowledge, belief, art, morals, law, custom, and any other abilities or habits that the individual has acquired as part of his social environment [61]. In recent years, the world has become a much smaller interactive arena due to the convergence of new media [62]. Social networks have enabled individuals to transcend geographical boundaries [63]. As a result, individuals with different cultural backgrounds are able to respect and understand each other's traditions and norms of living [64]. Social networks have provided an ideal platform for learning about different cultures. There are many posts about movies, music, local clothing and other cultural aspects of countries published daily on social networks such as Facebook, Twitter,

Instagram, and Telegram. It is common for people to promote handicrafts on social networking sites and facilitate the purchase and sale of these products. Therefore, it is necessary to provide laws supporting social networking exchanges. In addition, regulations must be created in order to ensure that the culture of a country is properly introduced in social networks, since these contents have both a positive and negative impact on people's judgments and perceptions of countries.

*3.8. Topic 8. Information Technology*

The advancement of technology has had a significant impact on the way we acquire knowledge and learn. In recent years, online technology has become increasingly important in the field of education and learning [65]. Using electronic resources to facilitate formal education is known as e-learning [66]. During the COVID-19, traditional educational methods were replaced by e-learning [67], which became the primary solution for schools and universities. Due to the widespread use of e-learning, there should be laws regarding plagiarism, copyright issues, and data protection.

Additionally, telehealth and telemedicine services were widely used during the COVID-19 pandemic. Patients and clinicians have been able to come together without the fear of contracting COVID-19 through telehealth services [68]. However, there are some concerns regarding reimbursement for services, patient privacy, and data sharing that require regulatory intervention. Having private information leaked affects the individual's personal life, leading to bullying, an increase in insurance premiums, as well as the loss of employment due to medical history. Hence, ensuring the security, privacy, and trustworthiness of information is important [69]. The lack of qualified personnel in the cybersecurity field makes it difficult for many organizations to adopt a standard approach or framework for cybersecurity [70]. Cyber threats must be protected by law in order for organizations to remain safe.

*3.9. Topic 9. Political*

A major issue in Iran in recent years has been the nuclear negotiations. A number of agreements have been signed in accordance with the negotiations, whose non-compliance has resulted in sanctions against Iran. Typically, sanctions involve restrictions on trade as a means of coercing another country to comply with an international agreement or norm of conduct [71]. In the absence of a decision from the United Nations Security Council, these measures which called countermeasures are taken against international wrongdoers [72]. There have been comprehensive unilateral economic sanctions imposed on Iran in recent years by groups of countries (not the United Nations). From 2012, they have intensified as a result of international uncertainty regarding Iran's nuclear program's peaceful purpose, as well as the inadequacy of the country's trust-building efforts [73]. Iran's free trade has been affected by these sanctions. Iran can benefit from partnerships with neighboring countries such as Iraq, Qatar, Oman, Pakistan, Tajikistan and

Uzbekistan to reduce the impact of sanctions. To conduct negotiations effectively, it is necessary to establish laws for better communication with other countries.

*3.10. Topic 10. Government*

It is essential for the government to have laws in order to govern. There are several entities within the government, including government boards, government companies, and government property. In order to run government affairs, people are employed by government companies. Although some of the work can be handled by these government forces and companies, due to the high volume of work, some work must be assigned to the private sector.

In recent years, privatization has gained popularity as a panacea for resolving the organizational problems of governments by reducing the role of the state and promoting private sector enterprise growth. As stated in the Iranian Constitution, No.144, public companies must become private and a rapid program of privatization was launched by the government, including major industries, banks, insurance companies, airlines, shipping companies, etc. [74]. Following the government's decision to privatize 80% of the state's assets in 2003, the privatization of the healthcare system rapidly increased [75] with the purpose of improving efficiency, social welfare, and system quality [76]. It is urgent that the government of Iran expands and develops its private sector, increases the efficiency and effectiveness of infrastructure and utility provision, and increases investment from both domestic and foreign sources. Privatization in Iran is intended to increase efficiency [74]. A privatization program usually begins with a period of partial privatization, when only non-controlling shares of firms are sold on a stock exchange [77]. This privatization will require legal requirements in order for monitoring and evaluation mechanisms to be effective, otherwise it will result in a reduction in the quality of services and social welfare. According to Akhavan's [76] study, privatization of the health care sector has resulted in a lower-quality, unfair, and more expensive health care system in Iran.

## 4. Discussion and conclusions

Analysis and organization of documents are made possible by text clustering using topic modeling. In this study, topic modeling was conducted on Iran's laws. In the first step, web scraping was used to extract the data and then preprocessing was applied. The next step involved converting the texts into vectors using TF-IDF, and clustering the data by LDA method.

The collected data contains a variety types of laws. In this study, the topic modeling was based solely on the regulations. Based on the results, regulations cover a wide range of subjects, which are divided into 10 topics, including Economic, Customs, Housing and Urban Development, Agriculture, Insurance, Legal and judicial, Cultural, Information Technology, Political, and Government.

There is a great deal of regulation related to the economy, which is the most important topic. The economic conditions of a country are among the most important issues, and economic development leads to prosperity in people's lives and high satisfaction with the conditions of their society. As a result, this topic requires special attention.

The next topic is customs, which controls exports and imports. There are various goods people require for their daily lives, some of which are essential goods, such as medicine, and should not be in short supply. Alternatively, some manufactured products that are in excess of the population's needs may be exported, resulting in economic prosperity and foreign exchange. Therefore, it is necessary to establish laws for customs and duties on goods that do not cause harm to people's health or daily lives.

Housing and Urban Development are another important topic. In centuries ago, humans realized that by living together, they were able to increase their productivity and also protect themselves better against their enemies. Aside from this, due to the prevalence of agriculture, it was difficult to move around. The increase in prosperity and security led to an increase in population and urbanization, which necessitated the creation of laws to govern these cities and guide their development.

Next, most of the regulations relate to agriculture. The effects of climate change and drought require that the necessary measures be taken to ensure the provision of food to the people and to prevent farmers' losses.

Insurance was created in human societies as a result of the need to feel secure and to ensure the safety of people and their property during disasters. Considering the wide range of insurance fields from automobile, health, tourism to life insurance and retirement, it is necessary to create laws to protect the interests of the insured and insurers.

Next topic is legal and judicial. In spite of the fact that peaceful living has been one of mankind's ideals and aspirations throughout history, conflict and differences have become inseparable parts of human life due to differences in behavior and opinions. Families, marriages, and divorces are often the subject of such disputes. Legislators must consider measures to address these issues.

There is a separate topic devoted to culture, which illustrates its importance. All cultures have their own values, beliefs, and norms, which make them unique. People are becoming familiar with different cultures due to the expansion of social networks. Consequently, countries see themselves at risk of cultural invasion. In order to resolve cultural issues, they create laws.

As information technology has affected all aspects of human life, legislation is necessary for education, security, privacy, and data protection on the Internet, social networks, fraud prevention, and many other issues. It is important to understand that these laws should not be restrictive and disrupt people's lives, but rather should aim to ensure the safe use of resources.

The political laws is another topic of this research. The political views of a country affect their interactions with other countries and their international relations. In order to protect the interests of the country and to establish friendly relations with other countries, political laws are necessary.

The last topic is the government; whose duty is to plan for the development of the country. Considering the importance of the government in the administration of the country and the impact of this institution on the economy as well as the broad powers it has, it is logical to have a separate issue for its regulations.

In this study, topic modeling has been conducted only on the regulations data. It is possible to consider other types of laws in future works. Further, the TF-IDF method was used for vectorization, which can be supplemented by other methods such as ParsBERT.